# A Fixed-Parameter Study on Propositional Dynamic Logic

Mohammad Javad Hosseinpour[i]

*Farzad Didehvar[ii]*

[i] Amirkabir University of Technology, m.j.hosseinpour@aut.ac.ir

[ii] Amirkabir University of Technology, didehvar@aut.ac.ir

*Abstract*

Since its establishment, propositional dynamic logic (PDL) has been a subject of intensive academic research and frequent use in the industry. We have studied the complexity of some PDL problems and in this paper, we show results for some special cases of PL and PDL.

## 1. Introduction

Dynamic logic is a family of logic systems for studying the behavior of programs and can be used to verify the properties of terminating programs. While dynamic logic and closely related logics can be used in a variety of fields and problems, including knowledge representation (De Giracomo & Lenzerini, 1994) and linguistics (Kracht, 1995), the high complexity of its satisfiability problem (EXPTIME-complete (Fischer, 1979)) may cause problems in the utilization of the power of dynamic logic.

### 1.1. PDL Syntax And Semantics

Syntax of PDL can be described by defining two sets: a set of formulas ($\Phi$) and a set of programs ($\Pi$). These sets are defined by the following rules (Harel, 2000):

Rules for formulas:

I) Every atomic formula is a formula.
II) 0 is a formula.
III) If $\varphi$ and $\psi$ are formulas, then $\neg \varphi$, $\varphi \wedge \psi$, and $\varphi \vee \psi$ are formulas.

Rules for programs:

IV) Every atomic program is a program.

V) If $\alpha$ and $\beta$ are programs, then $\alpha;\beta$, $\alpha \cup \beta$, and $\alpha^*$ are programs.

Rules for deriving programs from formulas and vice versa:

VI) If $\varphi$ is a formula and $\alpha$ is a program, then $[\alpha]\varphi$ is a formula.
VII) If $\varphi$ is a formula, then $\varphi?$ is a program.

The semantics of propositional dynamic logic can be defined via Kripke frames. A Kripke frame is the pair $\mathfrak{K} = (K, m_\mathfrak{K})$ where K is the set of states and $m_\mathfrak{K}$ is the meaning function from each atomic formula to a subset of K and from each atomic program to a subset of $K^2$. Intuitively, we can say that $m_\mathfrak{K}$ contains the states that satisfy a certain formula and all possible input/output pairs of a certain program. We can define the semantics of every formula and program in PDL using induction. If $\varphi$ and $\psi$ are formulas and $\alpha$ and $\beta$ are programs, then:

I) $m_\mathfrak{K}(0) \stackrel{\text{def}}{=} \emptyset$
II) $m_\mathfrak{K}(\varphi \to \psi) \stackrel{\text{def}}{=} (K - m_\mathfrak{K}(\varphi)) \cup m_\mathfrak{K}(\psi)$
III) $m_\mathfrak{K}(\alpha;\beta) \stackrel{\text{def}}{=} m_\mathfrak{K}(\alpha) \circ m_\mathfrak{K}(\beta)$
IV) $m_\mathfrak{K}(\alpha \cup \beta) \stackrel{\text{def}}{=} m_\mathfrak{K}(\alpha) \cup m_\mathfrak{K}(\beta)$
V) $m_\mathfrak{K}(\alpha^*) \stackrel{\text{def}}{=} \bigcup_{n \geq 0} m_\mathfrak{K}(a)^n$
VI) $m_\mathfrak{K}([\alpha]\varphi) \stackrel{\text{def}}{=} K - (m_\mathfrak{K}(\alpha) \circ (K - m_\mathfrak{K}(\varphi)))$
VII) $m_\mathfrak{K}(\varphi?) \stackrel{\text{def}}{=} \{(u,u) | u \in m_\mathfrak{K}(\varphi)\}$

## 1.2. Previous Works

A substantial amount of work on logics designed to verify the properties of programs can be contributed to Hoare's work and his logic, aptly named Hoare Logic (Hoare, 1969). His work was continued by Salwicki's Algorithmic Logic (Salwicki, 1977) and Pratt's Dynamic Logic (Pratt, 1976). In 1979, Fischer and Ladner introduced the propositional version of Dynamic Logic (Fischer, 1979). Since then, many others variants of PDL have been introduced. PDL with intersection (IPDL) (Lange & Lutz, 2005), PDL with parallel composition (RSPDL) (Balbiani & Tinchev, 2014), and PDL with converse (CPDL) (M, 1985) are just a few examples of these new logics.

The time complexity of the satisfaction problem for these logics has also been subject to study. Propositional dynamic logic is known to be EXPTIME-complete, while IPDL is 2EXPTIME-complete, CPDL is EXPTIME and RSPDL is undecidable.

Since the time complexity of different variants of PDL is high (mostly EXPTIME-complete and above), it is worthwhile to search for different techniques to reduce their time complexity. The technique used in this paper is trying to study special cases of PL and PDL using fixed-parameter algorithms. In this method, one of the parameters of the algorithm (like k) input is focused on, and the complexity of the algorithms is computed as a parameter of input size and k (Gurevich, Stockmeyer, & Vishkin, 1984). This paper uses this technique by restricting the number of Boolean negations that can appear in a formula.

## 2. Fixed-Parameter Results for PL And PDL

In this section, special cases of PL and PDL are studied. But before introducing these two algorithms, a few theorems will be stated and proved, to provide a foundation for the fixed-parameter algorithms.

***Definition 2.1.*** Shaefer's Dichotomy Theorem (SDT) (Shaefer, 1978) is a theorem that partitions Boolean satisfaction problems into two categories: $P$ and $NP$ (assuming $P \neq NP$). This theorem states that a Boolean satisfaction problem $F$ can be solved in polynomial time, if and only if it satisfies at least one of the six following conditions:

- I) $F$ is satisfied if all variables are false.
- II) $F$ is satisfied if all variables are true.
- III) $F$ is definable by a CNF formula where each conjunct has at most one negated variable.
- IV) $F$ is definable by a CNF formula where each conjunct has at most one unnegated variable.
- V) $F$ is definable by a CNF formula where each conjunct has at most two literals.
- VI) $F$ is the set of solutions of a system of linear equations over the two-element field $\{0, 1\}$.

Using this tool, we can prove our first theorem.

***Theorem 2.1.*** The satisfiability problem in positive propositional logic is PTIME.

***Proof.*** SDT's second condition.

***Definition 2.1.*** A valuation of a formula like $f$ is the set of positive variables in the valuation, and is shown as $T(f)$.

***Lemma 2.1.*** If $f$ is a formula in positive propositional logic, $T_1(f) \subseteq T_2(f)$ and $T_1(f)$ satisfies $f$, then $T_2(f)$ satisfies $f$.

***Proof.*** Induction on formula length. If $\alpha$ is an atomic formula, then $\{\alpha\}$ is the only satisfying valuation and therefore the lemma stands. Now, suppose $a$ and $b$ are formulas for which the lemma is true. There are only two ways to make a formula with $a$ and $b$ in positive propositional logic:

*Case 1.* $a \lor b$: If $T_1(a \lor b)$ satisfies $a \lor b$, then it either satisfies $a$ or $b$; we know that $T_2$ satisfies at least one of $a$ or $b$, so it satisfies $a \lor b$.

*Case 2.* $a \land b$: If $T_1(a \land b)$ satisfies $a \land b$, then it satisfies both $a$ and $b$. Then if $T_1(a \land b) \subseteq T_2(a \land b)$, based on the assumption, $T_2$ satisfies both $a$ and $b$, so it satisfies $a \land b$.

***Theorem 2.2.*** The satisfiability problem in propositional logic with one Boolean negation is PTIME.

*Proof.* If a Boolean formula is transformed to the equal CNF formula (in polynomial time), there can be four possible ways to put a Boolean negation operator in the formula:

*Case 1.* Before a variable.

By SDT's third condition, the problem belongs to $P$.

*Case 2.* Before a group of variables in a conjunction.

After using De Morgan's law and distributing positive literals into negative literals, a formula satisfying SDT's third condition will be obtained.

*Case 3.* Before a conjunction.

After applying De Morgan's law, each negative literal will be a separate conjunction, satisfying SDT's third condition.

*Case 4.* Before a group of conjunctions.

For this case, the following algorithms can be used. A proof of the correctness of this algorithm will also be provided.

***Algorithm 2.1.***

1- Apply De Morgan's law twice. A formula of the form $\bigwedge_i \bigvee_j x_{ij} \land \bigvee_m \bigwedge_n \sim y_{mn}$ will be reached.
2- For every conjunction $\bigwedge_n \sim y_n$:
   2-1- Define the valuation $T(f) = V_{total} - \{y | y \text{ is present in } \bigvee_n \sim y_n\}$.
   2-2- If $T(f)$ satisfies the formula, return $T(f)$ and exit.
3- Return *none*.

It is obvious that if *Algorithm 1* returns a valuation, the valuation satisfies the formula. To show that if the formula has at least a satisfying valuation, the algorithm returns one.

Suppose after testing valuations $T_1, \ldots, T_n$, no satisfying valuation has been found, but there is a valuation $T_{true}$ that satisfies the formula. There are only two possibilities for $T_{true}$:

Case 1. For every $T_i$, $T_{true} - T_i \neq \emptyset$: it is obvious that $T_{true}$ does not satisfy $\bigvee_m \bigwedge_n \sim y_{mn}$ and therefor does not satisfy the formula, reaching a contradiction.

Case 2. There is a $T_i$ such that $T_{true} \subseteq T_i$: we know that $T_i$ satisfies the negative part of the formula, and following *lemma 2.1*, it also satisfies the positive part of the formula, therefor satisfying the entire formula and reaching a contradiction.

***Theorem 2.3.*** The satisfiability problem with two Boolean negations is PTIME.

***Proof.*** All possible combinations of two Boolean negations can be reduced to either instances of formulas with one negation, or directly to one of the conditions of Shaefer's Dichotomy Theorem.

***Definition 2.2.*** A propositional formula is **AND-only** if it only contains atomic propositions, the Boolean AND operator and the Boolean NOT operator.

***Definition 2.3.*** A propositional formula is **OR-only** if it only contains atomic propositions, the Boolean OR operator and the Boolean NOT operator.

***Theorem 2.4.*** The satisfiability problem for AND-only formulas with a fixed number of Boolean negations is PTIME.

***Proof.*** After distributing the Boolean negations to atomic propositions, a formula with alternating AND and OR operators is achieved. This formula can be represented as a tree, with each leaf being an atomic proposition, and inner nodes being AND or OR operators. The height of this tree is at most $k$, and so the number of its leaves is at most $2^k$. Therefor, there are at most $2^k$ valuations for the formula and so the complexity of the satisfiability problem is $O(n2^k)$.

***Theorem 2.5.*** The satisfiability problem for OR-only formulas with a fixed number of Boolean negations is PTIME.

***Proof.*** The only non-trivial form of OR-only formulas has the form $\neg \bigvee_i \Phi_i$ that after distributing the negation, creates an AND-only formula. Since the number of Boolean negations is fixed, $i \leq k$ and therefore the problem is PTIME.

***Theorem 2.6.*** Every PDL formula has an equivalent formula with Boolean negations applied only on propositions.

***Proof.*** If a non-atomic formula such as $\psi$ is negated in the original formula, it can be written as $[\psi?](\sigma \wedge \neg \sigma)$ where $\sigma$ is an atomic formula, The two formulas are equivalent:

$$m_{\mathfrak{R}}([\psi?](\sigma \wedge \neg\sigma)) = m_{\mathfrak{R}}([\psi?]0) = K - \left(m_{\mathfrak{R}}(\psi?) \circ (K - m_{\mathfrak{R}}(0))\right)$$
$$= K - (m_{\mathfrak{R}}(\psi?) \circ K) = K - (\{(u,u) | u \in m_{\mathfrak{R}}(\psi)\} \circ K)$$
$$= K - \{u | u \in m_{\mathfrak{R}}(\psi)\} = K - m_{\mathfrak{R}}(\psi) = m_{\mathfrak{R}}(\neg\psi)$$

## 3. Conclusion

PL and PDL, due to their different properties, don't seem to react to fixing parameters similarly. Fixing the number of atomic formulas is such an example [n]. Providing fixed-parameter algorithms for PL is an easier task than doing so for PDL. However, since PL can be fully expressed in PDL, obtained results in PDL can almost always apply to PL, making these results much stronger.